\documentclass{ws-procs9x6-cpt16}
\begin{document}

\newcommand{\refeq}[1]{(\ref{#1})}
\def\etal {{\it et al.}}


\title{Prospects for Testing Lorentz and CPT Invariance\\ 
in the Top-Quark Sector}

\author{Z.\ Liu}

\address{Physics Department, Indiana University\\
Bloomington, IN 47405, USA}

\begin{abstract}
We present how to further the search for Lorentz and CPT violation in the top-quark sector after the first measurement in this sector by D0. We compute the Lorentz-violating matrix element for top pair production via gluon fusion, which allows a similar but more precise measurement at the LHC. For CPT violation, we analyze single-top production, derive the modified matrix elements, and predict possible signals.
\end{abstract}

\bodymatter

\section{The top-quark sector in the SME}

Many experiments involving the top quark have been performed to determine the top-quark properties and test new physics. With the advent of the LHC, it is also possible to test some fundamental symmetries, for example Lorentz and CPT symmetries, using the top quark, given the great statistical power of the LHC data. The use of the top quark is motivated by its large mass, since the Lorentz and CPT violation may be largest for the top quark, and by its short lifetime, since it decays before hadronization and can be treated as a free quark.\cite{Berger:2015yha}

The Lorentz and CPT violation can be described by an effective field theory called the Standard-Model Extension (SME).\cite{Colladay:1996iz,Colladay:1998fq} Numerous experiments in various fields of physics have been conducted to search for Lorentz and CPT violation.\cite{datatables} The only experiment testing Lorentz symmetry in the top-quark sector of the SME has been carried out by the D0 collaboration, with the dominant production process being $q\overline{q}\rightarrow t\overline{t}$ and the CPT-even coefficients measured to about 10\%.\cite{Abazov:2012iu} To make a similar measurement at the LHC, we need a theoretical analysis of the $t\overline t$ production via gluon fusion ($gg\rightarrow t \overline t$), the dominant $t\overline t$ production process at the LHC. We expect these coefficients to be measured to a few percent, for the LHC has much higher statistical power. On the other hand, our analysis of single-top production lays the foundations for measuring CPT violation in the top-quark sector.\cite{Berger:2015yha}

To search for Lorentz and CPT violation, we can look for sidereal signals. For CPT-odd effects, we may also use a different kind of signal called an asymmetry,
\begin{equation}
{\mathcal{A}_{\mathrm{CPT}}}\equiv\dfrac{R-\overline{R}}{R+\overline{R}},
\label{zliu}
\end{equation}
where $R$ and $\overline{R}$ are the rates of one process and its CPT-conjugated process.

\section{Top-antitop pair production}

Top quarks are produced dominantly in pairs via quark fusion and gluon fusion in hadron colliders. The top and antitop then decay. The squared matrix element for the whole process can be written as the product of production and decay parts in the narrow-width approximation.

We start with the Lagrange density in the top-quark sector of the minimal SME assuming the only nonzero coefficients are those that involve the top-quark fields. For the matrix element we calculate, all physically observable effects come from the symmetric part of $c_{\mu\nu}$ by field redefinitions. Moreover, CPT violation is unobservable in $t\overline t$ production and decay at the leading order.\cite{Berger:2015yha}

The leading-order SME corrections to the matrix element for $gg\rightarrow t \overline t$, which is the dominant $t\overline t$ production process at the LHC, can be calculated as follows. Extracting from the SME Lagrange density, we obtain the modified Feynman rules, which include insertions on the quark-gluon vertices and the top-quark propagators. To compute the contribution from the vertex insertions, we add five diagrams, each of which has one vertex insertion, to the Standard Model (SM) $s$, $t$ and $u$ channel tree-level diagrams, take the modulus square, average over polarizations and colors, and sum over spins. For the corrections from the propagators, we use the full propagator, which can be obtained from the modified momentum-space Dirac equation. The sum of all these SME corrections is symmetric under $\mu \leftrightarrow \nu$, which is consistent with the previous discussion that the antisymmetric part of $c_{\mu\nu}$ is not physically observable.\cite{Berger:2015yha}

The combination of the production and decay parts\cite{Berger:2015yha} leads to the matrix element for the whole process, which can then be used to obtain experimental signals like the cross sections and sidereal variations.

\section{Single-top production}

Although CPT violating effects are absent in $t\overline t$ production, they appear in single-top production, which includes the $s$ channel ($q\overline{q'}\rightarrow t\overline{b}$), $t$ channel ($bq\rightarrow tq'$ and $b\overline{q}\rightarrow t\overline{q'}$) and $tW$ mode ({$bg\rightarrow tW^-$}). 

For the Lagrange density, in addition to the assumptions we make in the $t\overline t$ production analysis, we further assume the only nonzero coefficient is $b_\mu$. This leads to an insertion on a top-quark line and modified spin sums for top quarks. To find the spin sums, we use Appendix A of Ref.\ \refcite{Colladay:1996iz} to obtain the approximate solutions to the modified Dirac equation, compute the spin sums in the zero-momentum frame of the particle, and finally do an observer Lorentz transformation.

In the narrow-width approximation, the SME corrections to the matrix elements for all four production processes can be calculated in a similar way. The SME corrections to the corresponding single-antitop production processes have the same magnitudes as those single-top processes, but have opposite signs. The decay part is found to be the same as the SM results.\cite{Berger:2015yha}

\section{Signals}
For $t\overline t$ production, a similar sidereal analysis at the LHC is expected to measure $c_{\mu\nu}$ to about a few percent. For single-top production, in addition to sidereal variations, the asymmetry defined in Eq.\ \refeq{zliu} gives another type of signal. For example, the cross sections of $tW^-$ and $\overline{t}W^+$ modes are the same in the SM\cite{Bernreuther:2008ju} but have opposite SME corrections, and this asymmetry is sensitive to $b_Z$ and $b_T$. The estimated sensitivity to $b\cdot p/s$ is about 5\%.

\section*{Acknowledgments}
I am very grateful to M.S. Berger and V.A. Kosteleck\'y for their collaboration. This work is supported partly by DOE grant DE-SC0010120 and by the Indiana University Center for Spacetime Symmetries (IUCSS).

\end{document}